\documentclass[aps,prd,preprint, onecolumn, tightenlines, notitlepage, superscriptaddress, nofootinbib, preprintnumbers, floatfix]{revtex4-2}

\usepackage{amstext}
\usepackage{amssymb}
\usepackage{amsmath}
\usepackage{graphicx}
\graphicspath{{plots/}}
\usepackage{hyperref}
\usepackage{url}
\usepackage{color}
\usepackage{ulem}
\usepackage[utf8]{inputenc}
\pdfoutput=1
\usepackage[x11names]{xcolor}
\usepackage{textcomp}

\usepackage{epsfig,amsfonts,mathrsfs,amsmath,amssymb,graphicx,color,slashed,multirow}
\usepackage{amsmath,latexsym,amssymb,graphicx,slashed,hyperref,color,enumerate,url,cancel,gensymb}
\usepackage{textcomp}

\hypersetup{colorlinks,citecolor= nicegreen,linkcolor= nicegreen}
\definecolor{nicered}{rgb}{0.7,0.1,0.1}
\definecolor{nicegreen}{rgb}{0.1,0.5,0.1}

\def\beq{\begin{equation}}
\def\eeq{\end{equation}}
\def\beqn{\begin{eqnarray}}
\def\eeqn{\end{eqnarray}}

\definecolor{darkblue}{rgb}{0,0,80}

\newcommand{\diag}{\operatorname{diag}}

\AtBeginDocument{\hypersetup{citecolor=Green3,linkcolor=Blue1,urlcolor=Blue1}}

\begin{document}

\title{{\Large Short-baseline oscillation scenarios at JUNO and TAO }}

\author{V.~S.~Basto-Gonzalez}\email{victor.basto@unipamplona.edu.co}
\affiliation{Universidad de Pamplona, km 1, v\'{i}a salida a Bucaramanga, Campus Universitario, 543050 Pamplona, Colombia}
\affiliation{Departamento de F\'{i}sica, Pontif\'{i}cia Universidade Cat\'{o}lica do Rio de Janeiro,
C. P. 38071, 22452-970, Rio de Janeiro, RJ, Brazil}

\author{D.~V.~Forero}\email{dvanegas@udemedellin.edu.co}
\affiliation{Universidad de Medell\'{i}n, Carrera 87 N° 30 - 65 Medell\'{i}n, Colombia}

\author{C.~Giunti}
\email{carlo.giunti@to.infn.it}
\affiliation{Istituto Nazionale di Fisica Nucleare (INFN), Sezione di Torino, Via P. Giuria 1, I--10125 Torino, Italy}

\author{A.~A.~Quiroga}\email{alarquis@puc-rio.br}
\affiliation{Departamento de F\'{i}sica, Pontif\'{i}cia Universidade Cat\'{o}lica do Rio de Janeiro,
C. P. 38071, 22452-970, Rio de Janeiro, RJ, Brazil}

\author{C.~A.~Ternes}\email{ternes@to.infn.it}
\affiliation{Istituto Nazionale di Fisica Nucleare (INFN), Sezione di Torino, Via P. Giuria 1, I--10125 Torino, Italy}

\begin{abstract}
We study the sensitivity of JUNO and TAO to the oscillations induced by two well-motivated scenarios beyond the standard model: Large Extra Dimensions (LED) and light sterile neutrinos in the context of 3+1 neutrino mixing.
We find that JUNO+TAO can set competitive bounds on the parameter space of each scenario. In particular, we find that JUNO+TAO can be competitive with MINOS, DUNE or KATRIN in the context of LED. If LED are present in nature, we show that the parameters could be measured with a similar precision as the standard oscillation parameters. We also show that JUNO+TAO can test nearly all of the parameter space preferred by Gallium experiments in the context of 3+1 mixing. Finally, we discuss the possibility to distinguish the two scenarios from each other.

\end{abstract}

\maketitle
\newpage
\tableofcontents

\section{Introduction}
\label{sec:intro}

Neutrino oscillations is the leading mechanism that successfully explains neutrino flavor conversion. Neutrino oscillations have been observed by experiments measuring neutrinos from the Sun, Earth's atmosphere, nuclear reactors, and particle accelerators. The existence of neutrino oscillations implies that neutrinos are massive and mixed and provides a direct evidence of physics beyond the standard model.

Currently, the best scenario that fits most of the observed neutrino oscillation data requires three active neutrinos, with at least two of them having different non-zero masses~\cite{deSalas:2020pgw,Esteban:2020cvm,Capozzi:2021fjo}. It should be noted that neutrino oscillation experiments are not sensitive to the absolute neutrino mass scale, which is expected to be measured by different means, i.e. using decay experiments or cosmological observations
(see Refs.~\cite{DeSalas:2018rby,Formaggio:2021nfz} for recent reviews on this topic). Some of the parameters describing neutrino oscillations are fairly well measured ($\Delta m^2_{21}, |\Delta m^2_{31}|, \sin^2\theta_{12}, \sin^2\theta_{13}$), while the measurement of others needs further improvement ($\delta_{CP}, \sin^2\theta_{23}$, the sign of $\Delta m^2_{31}$). The phase that encodes the possibility of leptons violating the Charge-Parity~(CP) symmetry (the so call Dirac CP phase) has been measured by T2K~\cite{T2K:2019bcf} and NOvA~\cite{NOvA:2021nfi}. However, there is some disagreement in the results, as can be seen in Fig. 6 of Ref.~\cite{NOvA:2021nfi}. It is also not clear if the true value of the atmospheric angle $\theta_{23}$ lies in the first or second octant, or if it is exactly maximal. The last unknown is the neutrino mass ordering. From the current data it is not clear if $\Delta m_{31}^2 > 0$ or $\Delta m_{31}^2 < 0$.

Future neutrino oscillation experiments such as DUNE~\cite{DUNE:2020ypp}, Hyper-Kamiokande~\cite{Hyper-Kamiokande:2018ofw}, ORCA~\cite{KM3Net:2016zxf}, and JUNO~\cite{JUNO:2015sjr,JUNO:2020ijm,JUNO:2021vlw}, are expected to improve the current precision of the neutrino oscillation parameters and address the remaining unknowns discussed before. To fulfill these goals, large statistics and an unprecedented control of systematic uncertainties have to be achieved. The expected precision provides a unique avenue to probe different beyond the standard model scenarios that might produce sub-leading effects on top of the well established three-neutrino oscillation pattern.

Here, we focus on the next generation reactor experiment JUNO (the Jiangmen Underground Neutrino Observatory) and its near detector TAO (the Taishan Antineutrino Observatory). JUNO in combination with TAO aims to measure very precisely several of the neutrino oscillation parameters ($\sin^2(\theta_{12}), \Delta m^2_{21}, |\Delta m^2_{31}|$) and the neutrino mass ordering
(see Refs.~\cite{Zhan:2008id,Ge:2012wj,Capozzi:2013psa,Li:2013zyd,JUNO:2015sjr,Capozzi:2015bpa,Capozzi:2020cxm,Forero:2021lax}). Apart from this, a very rich program of new physics searches will be possible using JUNO
(see Refs.~\cite{Abrahao:2015rba,JUNO:2015sjr,Khan:2013hva,Ohlsson:2013nna,Bakhti:2014pva,Li:2014rya,Liao:2017awz,Li:2018jgd,Anamiati:2019maf,Porto-Silva:2020gma,deGouvea:2020hfl,JUNO:2020ijm,Cheng:2020jje,Martinez-Mirave:2021cvh,Delgado:2021vha}). In this paper we discuss
the sensitivity of JUNO and TAO to oscillation effects induced in two models of physics beyond standard three-neutrino mixing: Large Extra Dimensions (LED) and a light sterile neutrino (3+1).
In particular, we discuss the sensitivity of JUNO and TAO to the parameters of the two models.

The paper is structured as follows: in Section~\ref{sec:formal} we introduce the formalism for LED and 3+1 oscillations. In Section~\ref{sec:sim} we discuss the details of our simulations. Our main results are presented in Section~\ref{sec:res}. Finally, we summarize and draw our conclusions in Section~\ref{sec:conc}.

\section{Formalism}
\label{sec:formal}
In this section we discuss the relevant formalisms for LED and 3+1 neutrino oscillations. The results of this section serve as physics input in our numerical analyses discussed in the next sections.

\subsection{Neutrino oscillations in presence of Large Extra Dimensions}

The Kaluza-Klein (KK) theory, in the contexts of Large Extra Dimensions, was introduced in order to solve the hierarchy problem between the electroweak and the Planck scale~\cite{Arkani-Hamed:1998jmv,Arkani-Hamed:1998sfv,Antoniadis:1998ig}. 
LED models incorporate neutrino masses~\cite{Dienes:1998sb,Arkani-Hamed:1998wuz} by allowing standard model gauge singlets, such as gravitons or right-handed neutrinos, to propagate in a higher-dimensional bulk of extra dimensions while restricting the Standard Model (SM) particles to propagate in our four-dimensional brane. In these models, neutrinos acquire a Dirac mass term, which is strongly suppressed by the volume of the Large Extra Dimensions. Thus, the same mechanism which explains the gap between the electroweak and Planck scales generates naturally small neutrino masses. 

In the LED model, the SM is extended by adding three five-dimensional\footnote{It is assumed that the compactification manifold is asymmetric, i.e., that one of the extra dimensions is much larger than the sizes of the other dimensions, and is compactified on a circle of radius $R_{ED}$~\cite{Davoudiasl:2002fq}. The effective treatment is therefore five-dimensional.} fermions $\Psi^\alpha = (\psi_L^\alpha,\psi_R^\alpha)$, each associated to one active neutrino with flavor $\alpha$. From our four-dimensional point of view, these fields can be expressed as infinite towers of Kaluza-Klein modes, $\psi^{\alpha(n)}_{L,R}$, for $n = 0, \pm 1, \pm 2, \ldots$ Redefining the fields as
\begin{eqnarray}
    \nu_R^{\alpha (0)} &=& \psi_R^{\alpha (0)}\,,\nonumber\\
    \nu_{L,R}^{\alpha (n)} &=& \frac{1}{\sqrt{2}}\left(\psi_{L,R}^{\alpha (n)}+\psi_{L,R}^{\alpha (-n)}\right),
    \quad
    \text{for}
    \quad
    n = 1,\ldots,\infty\,,
\end{eqnarray}
the mass term in the Lagrangian can be written as~\cite{Dienes:1998sb,Barbieri:2000mg,Arkani-Hamed:1998wuz,Davoudiasl:2002fq}
\begin{eqnarray}
\mathcal{L}=m^D_{\alpha\beta}\left(\overline{\nu}_R^{\alpha(0)}\nu_L^{\beta}+\sqrt{2}\sum_{n=0}^\infty\overline{\nu}_R^{\alpha(n)}\nu_L^{\beta}\right) + \sum_{n=1}^\infty\frac{n}{R_{ED}}\overline{\nu}_R^{\alpha(n)}\nu_L^{\alpha(n)} + h.c. \,,
\end{eqnarray}
where $m^D_{\alpha\beta}$ are the entries of the Dirac mass matrix and $R_{ED}$ is the compactification  radius of the extra dimensions. Note that repeated indices are summed over. The diagonalization of this mass term can be performed in two steps~\cite{Davoudiasl:2002fq}. First diagonalize the Dirac mass term using $\diag(m_1^D,m_2^D,m_3^D) = R'{}^\dagger m^D U$ with the redefinitions
\begin{eqnarray}
\nu_L^{\alpha} &=& \sum_{i=1}^3 U_{\alpha i } \nu_L^{i (0)}\,,\nonumber\\
\nu_R^{\alpha(0)} &=& \sum_{i=1}^3 R'_{\alpha i } \nu_R^{i (0)}\,,\nonumber\\     
\nu_{L,R}^{\alpha(n)} &=& \sum_{i=1}^3 R'_{\alpha i } \nu_{L,R}^{i (n)}, \quad n=1,2,\ldots, \infty\,.   
\end{eqnarray}
The mass term can be written now as 
\begin{equation}
\mathcal{L}=\sum_{i=1}^3 \overline{\nu}_R^i M_i\nu_L^i + h.c. \,,
\end{equation}
where
\begin{equation}
        \nu_{L, R}^i=(\nu_{L,R}^{i (0)}, \nu_{L,R}^{i (1)}, \ldots)^T ,\qquad M_i = 
\begin{pmatrix}
m_i^D         & 0   & 0   & \ldots\\
\sqrt{2}m_i^D & 1/R_{ED} & 0   & \ldots\\
\sqrt{2}m_i^D & 0   & 2/R_{ED} & \ldots\\
\vdots  & \vdots & \vdots & \ddots
\end{pmatrix}
\end{equation}
are infinite-dimensional vectors and matrices. In a last step we diagonalize the $M_i$ using infinite-dimensional matrices $L_i$ and $R_i$. The mass states are then given by $\nu_L^i{}' = L_i^\dagger \nu_L^i$ and $\nu_R^i{}' = R_i^\dagger \nu_R^i$. Here we are interested in the decomposition of the standard flavor neutrinos $\nu_L^\alpha$, which becomes now 
\begin{equation}
    \nu_L^\alpha = \sum_{i=1}^3 U_{\alpha i}\sum_{n=0}^\infty L_i^{0 n}\nu_L^{i(n)}{}'\,,
\end{equation}
where the relevant entries of $L_i$ are given by~\cite{Dienes:1998sb,Dvali:1999cn,Mohapatra:2000wn}
\begin{equation}
    \left(L_i^{0n}\right)^2 = \frac{2}{1+\pi^2\left(m_i^D R_{ED}\right)^2+\left(\lambda_i^{(n)}\right)^2/\left(m_i^D R_{ED}\right)^2}\,,
    \label{eq:L_entrees}
\end{equation}
with $n=0,1,2, \ldots, \infty$. In this equation, the $\lambda_i^{(n)}$ are the eigenvalues of $R_{ED}^2M_iM_i^\dagger$ and can be obtained from solving the equation 
\begin{equation}
    \lambda_i^{(n)}-\pi\left(m_i^D R_{ED}\right)^2 \cot\left(\pi \lambda_i^{(n)} \right) = 0\,.
    \label{eq:lambda}
\end{equation}
The neutrino oscillation probability is then given by
\begin{equation}
    P(\nu_\alpha\rightarrow \nu_\beta) = \left\vert\sum_{i=1}^3\sum_{n=0}^\infty U_{\alpha i}U_{\beta i}^* \left(L_i^{0n}\right)^2\exp\left(-i\frac{\left(\lambda_i^{(n)}\right)^2L}{2ER_{ED}^2}\right) \right\vert^2 \,,
    \label{eq:oscprob}
\end{equation}
where we used $\nu_\alpha$ as abbreviation for the mostly active neutrino $\nu_L^\alpha$. Here, $E$ and $L$ are the neutrino energy and the distance traveled by the neutrino. From this formula it becomes clear that the standard mass splittings, characterizing the solar and atmospheric oscillations are given by~\cite{Basto-Gonzalez:2012nel,Berryman:2016szd,Carena:2017qhd,Stenico:2018jpl}
\begin{equation}
    \Delta m_{j1}^2 = \frac{\left(\lambda_j^{0}\right)^2-\left(\lambda_1^{0}\right)^2}{R_{ED}^2}\,.
    \label{eq:masssplittings}
\end{equation}
It is important to note that 
\begin{equation}
    \Delta m_{j1}^2 \neq \left(m_j^D\right)^2 - \left(m_1^D\right)^2\,.
\end{equation}
If this was the case, the LED parameters would also affect the main oscillations. Instead, the LED parameters can only induce new oscillation lengths on top of the regular $1-3$ and $1-2$ oscillations.
Note that this method limits the number of effective LED parameters to 2, the compactification radius $R_{ED}$ and the lightest Dirac mass term $m_0 = m_1^D (m_3^D)$ for normal (inverted) neutrino mass ordering. For a given $R_{ED}$ and $m_0$ we use Eq.~\eqref{eq:lambda} to calculate $\lambda_{1}^{(0)} (\lambda_{3}^{(0)})$ for normal (inverted) ordering. Next, we use Eq.~\eqref{eq:masssplittings} to calculate the remaining two $\lambda_{i}^{(0)}$, and then, substituting them into Eq.~\eqref{eq:lambda}, we get the other two $m_i^D$. With all three $m_i^D$ given, we obtain each $\lambda_{i}^{(n)}$ with $n>1$ by solving Eq.~\eqref{eq:masssplittings} (see Ref.~\cite{Basto-Gonzalez:2012nel} for more details on this procedure). Note that even though the sum  is infinite, we have checked that taking $n$ up to $n_{\text{max} }=4$ is enough for obtaining a convergent behavior for the experiments considered in this analysis.

Neutrino oscillations with Large Extra Dimensions have been studied in the context of solar~\cite{Davoudiasl:2002fq}, atmospheric~\cite{Esmaili:2014esa}, accelerator~\cite{MINOS:2016vvv,Berryman:2016szd,DiIura:2014csa,Carena:2017qhd,Stenico:2018jpl,DUNE:2020fgq} and reactor neutrinos~\cite{Girardi:2014gna,Machado:2011jt,Machado:2011kt}.
Interestingly, in Ref.~\cite{Machado:2011kt} the authors showed that the reactor and Gallium anomalies can be resolved with LED-induced oscillations. We will come back to this in the next section.

\subsection{Neutrino oscillations in presence of a light sterile neutrino}

Light sterile neutrinos have been a hot topic in neutrino physics since many years. Their presence has been motivated by several anomalies which appeared at short-baseline neutrino oscillation experiments. 

Light sterile neutrinos at the eV mass scale might explain the anomalous appearance of electron neutrinos and antineutrinos in LSND~\cite{LSND:2001aii} and MiniBooNE~\cite{MiniBooNE:2018esg,MiniBooNE:2020pnu},
although the recent results of the MicroBooNE experiment~\cite{MicroBooNE:2021sne,MicroBooNE:2021jwr,MicroBooNE:2021rmx}
disfavor the $\nu_\mu\to\nu_e$ interpretation of the MiniBooNE low-energy anomaly
(see, however, the caveats discussed in Ref.~\cite{Arguelles:2021meu}).
It is also interesting that the MicroBooNE data may be interpreted as
an indication of short-baseline $\nu_e$ disappearance~\cite{Denton:2021czb}.

The deficit obtained in Gallium experiments~\cite{Abdurashitov:2005tb,Laveder:2007zz,Giunti:2006bj,Barinov:2021asz} might be due to light sterile neutrinos. Note, however, that the values of the mixing parameters required to explain this anomaly are in tension with several other bounds
(see the discussions in Refs.~\cite{Barinov:2021mjj,Giunti:2021kab,Berryman:2021yan}). 

The deficit observed in the rates of reactor experiments, the reactor antineutrino anomaly~\cite{Mention:2011rk}, could be explained with light sterile neutrinos, too. Note, however, that this anomaly might also be due to a misunderstanding of the reactor neutrino fluxes~\cite{Berryman:2020agd,Giunti:2021kab}.

Recently, a strong indication in favor of such oscillations
has been claimed by the Neutrino-4
collaboration~\cite{NEUTRINO-4:2018huq,Serebrov:2020rhy,Serebrov:2020kmd}.
However, this result is controversial and has been criticized in
Refs.~\cite{PROSPECT:2020raz,Danilov:2020rax,Giunti:2021iti}.

As one can see, there is a lot of interest in light sterile neutrinos. 
For recent reviews we refer the reader to Refs.~\cite{Giunti:2019aiy,Diaz:2019fwt,Boser:2019rta,Dasgupta:2021ies}.
In the 3+1 framework it is assumed that
there is a non-standard massive neutrino $\nu_{4}$
with mass $m_{4} \gtrsim 1 \, \text{eV}$
such that
$ m_{1}, m_{2}, m_{3} \ll m_{4} $,
where $ m_{1}, m_{2}, m_{3} $ are the masses of the three standard massive neutrinos
$ \nu_{1}, \nu_{2}, \nu_{3} $.
In the flavor basis,
besides the three standard active neutrinos
$ \nu_{e}, \nu_{\mu}, \nu_{\tau} $
there is a sterile neutrino $\nu_{s}$,
and the mixing is given by
\begin{equation}
\nu_{\alpha}
=
\sum_{j=1}^{4} U_{\alpha j} \nu_{j}
\qquad
(\alpha=e,\mu,\tau,s)
,
\label{mixing}
\end{equation}
where $U$ is the $4\times4$ unitary mixing matrix.
The effective short-baseline (SBL) survival probability
of electron neutrinos and antineutrinos is given by
\begin{equation}
P_{ee}^{\text{SBL}}
=
1 - \sin^2\!2\theta_{ee} \, \sin^2\!\left( \dfrac{\Delta{m}^2_{41} L}{4 E} \right)
,
\label{Pee}
\end{equation}
where
$\Delta{m}^2_{41}=m_{4}^2-m_{1}^2$
and
$\sin^2\!2\theta_{ee} = 4 |U_{e4}|^2 ( 1 - |U_{e4}|^2 )$,
where $\theta_{ee}$ is an effective mixing angle
that coincides with $\theta_{14}$
in the standard parameterization of the
mixing matrix. At longer baselines the oscillations due to $\Delta m_{41}^2$ are washed out and the neutrino oscillation probability is modified to 

\begin{equation}
P_{ee}^{\text{LBL}}
=
(1 - |U_{e4}|^2)^2 P_{ee}^{3\times 3} + |U_{e4}|^4
,
\label{Pee_LBL}
\end{equation}
where $P_{ee}^{3\times 3}$ is the standard three-neutrino oscillation probability.

\section{Details of analysis}
\label{sec:sim}
In this section we describe the simulation and analysis details used in this paper. We basically follow the descriptions of Refs.~\cite{JUNO:2015zny,IceCube-Gen2:2019fet,JUNO:2021vlw} and~\cite{JUNO:2020ijm,JUNO:2021vlw} for the far and near detector, respectively. Throughout the paper we assume an exposure time of eight effective years (2400 days) at both detectors\footnote{An exposure of eight years with eight reactors is approximately equivalent to the exposure of six years with ten reactors at the far detector, previously used in many JUNO sensitivity analyses.}. 
\subsection{Far detector}

We assume the eight reactors configuration from Ref.~\cite{IceCube-Gen2:2019fet}, since two of the originally considered Taishan cores might only become operative in a later stage of the experiment~\cite{JUNO:2020ijm}.
Therefore, the main signal at JUNO will come from the two 4.6 GW reactor cores at the Taishan power plant and the six 2.9 GW reactors at the Yangjiang power plant.
We also include contributions from Daya Bay and Huizhou, at 215~km and 265~km distance. We simplify both contributions as  single reactors with 17.4 GW thermal power~\cite{JUNO:2015zny}.

The fluxes for each reactor core are parameterized as in Refs.~\cite{Mueller:2011nm,Huber:2011wv}. The fission fractions~\cite{JUNO:2020ijm} are assumed to be the same for all reactors and are kept fixed at: $f_{235} = 0.561$, $f_{238} = 0.076$, $f_{239} = 0.307$, $f_{241} = 0.056$.

Regarding the detector, we assume a fiducial mass of 20~kt and 82\% selection efficiency~\cite{JUNO:2021vlw}. The cross section is taken from Ref.~\cite{Vogel:1999zy}. Using this criteria we expect 49~(5) inverse beta decay (IBD) events per day from the main reactors (Daya Bay and Huizhou). For the analysis we also include an energy resolution of 3\%. 
Apart from the background contributions from Daya Bay and Huizhou which depend on the oscillation parameters, we also include accidental, fast neutron $^9$Li/$^8$He, $\alpha-$n and geo-neutrino background components, which we have extracted from Refs.~\cite{JUNO:2015zny,JUNO:2021vlw} and normalized to an exposure of 8 years.
Hence, the number of signal-events in the far detector is computed in the following way. The probability-weighted flux is given by 

\begin{equation}
    \Phi_\text{osc}(E) = \sum_\text{cores} \frac{W_\text{th,core}}{L_\text{core}^2}P_{ee}(E,L_\text{core})\Phi_0(E)\,,
\end{equation}
where $E$ is the neutrino energy, $W_\text{th}$ is the thermal power of the core, $L_\text{core}$ is the core-detector distance and $\Phi_0(E)$ is the flux from the reactor, where the contribution from each isotope is weighted with the above stated fission fractions and $P_{ee}(E,L)$ is the $\bar{\nu}_e$ survival probability. The number of signal events in the energy bin $k$ is then

\begin{equation}
    N_k = \mathcal{N}\int dE \int_{E_{k,\text{min}}}^{E_{k,\text{max}}} dE' \Phi_\text{osc}(E)\sigma_\text{IBD}(E)R(E,E')\,
\end{equation}
where the normalization constant $\mathcal{N}$ contains efficiencies, detector size and exposure time, $\sigma_\text{IBD}(E)$ is the cross section, $E'$ the reconstructed energy and 

\begin{equation}
    R(E,E') = \frac{1}{\sqrt{2\pi}\sigma_E}\exp\left(-\frac{\left(E-E'\right)^2}{2\sigma_E^2}\right)
\end{equation}
is used for energy smearing with the resolution given by

\begin{equation}
    \sigma_E = \alpha\sqrt{E/\text{MeV}-0.78}~\text{MeV}
\end{equation}
with $\alpha = 3\%$. The total number of events is obtained by adding signal and background contributions.

\subsection{Near detector}
The computation of events at the near detector is similar, but with a few important differences. We assume the signal from all reactors except the nearest one to be negligible. This is a valid simplification, because the contribution from the second Taishan reactor is supposed to be at the $\mathcal{O}(1\%)$ level~\cite{JUNO:2020ijm}, while the contribution from other backgrounds is at the $\mathcal{O}(10\%)$ level. 
Due to the short baseline, we cannot assume pointlike source and detector anymore, but we must average over the dimensions of the reactor (which is larger than the detector).
We assume a fiducial mass of 1~t and an energy resolution of 1.5\% following Refs.~\cite{JUNO:2020ijm,JUNO:2021vlw}.
We assume the near detector to be positioned at 30~m from the core and expect to observe 2700 IBD events per day in the absence of short baseline oscillations.
We include accidental, fast neutron and $^9$Li/$^8$He background components. The spectral forms are extracted from Ref.~\cite{JUNO:2015zny}, but normalized to produce 190, 200 and 54 events per day~\cite{JUNO:2020ijm} for the three background components. The background/signal ratio that we obtain with these parameters is $\sim 16\%$ in the absence of short baseline oscillations.
Due to the vertexing capabilities of TAO, and for the purpose of this analysis we consider the near detector
to be divided in two segments with fiducial masses of  0.5~t.
The spherical detector is split in half along the direction of propagation with the two average baselines given by $(30\pm 3/8 \times 0.65)$~m~\cite{Berryman:2021xsi}.

\subsection{Statistical analysis}
\label{sec:Statistical}

In our analyses, we use a standard $\chi^2$ test hypothesis to quantify our results. For a given set of oscillation parameters $\vec{p}$ it is given by

\begin{equation}
    \chi^2_D(\vec{p},\vec{\alpha}) = \sum_{i=1}^{200}\frac{(N^D_i - T^D_i(\vec{p},\vec{\alpha}))^2}{N^D_i}\,,
\end{equation}
where $D=N_1,N_2$ for the near detector segments and $D=F$ for the far detector. Here, $N^D_i$ is the fake data spectrum and $T^D_i$ the prediction for the oscillation parameters $\vec{p}$.
It also depends on the systematic uncertainties $\vec{\alpha}$. 
In both quantities the background components are already included. 
In our analyses, we include several sources of systematic uncertainties which are penalized by 
\begin{equation}
  \chi^2_{\text{sys.}}(\vec{\alpha})  = \sum_k \frac{(\alpha_k - \mu_k)^2}{\sigma_k^{2}}\,,
\end{equation}
where $\mu_k$ and $\sigma_k$ are the expectation values and standard deviations for
each systematic uncertainty $\alpha_k$.
The uncertainties that we consider are: 
\begin{itemize}
    \item A 0.8\% normalization of the thermal power of each reactor.
    \item A 2\% overall normalization of the flux.
    \item A 0.13\% normalization of the selection efficiency.
    \item A shape uncertainty in each bin. We will vary the penalty on this quantity.
\end{itemize} 
Flux-related uncertainties are correlated among near and far detector, while the selection efficiencies are not. Therefore, the whole $\chi^2$ function is given by 

\begin{equation}
    \chi^2(\vec{p}) = \min_{\vec{\alpha}}\left\{\chi^2_{N_1}(\vec{p},\vec{\alpha})+\chi^2_{N_2}(\vec{p},\vec{\alpha})+\chi^2_F(\vec{p},\vec{\alpha})+\chi^2_{\text{sys.}}(\vec{\alpha})\right\}\,.
    \label{eq:chi2}
\end{equation}
It has been shown~\cite{Girardi:2014gna} that the $\theta_{13}$ measurement is fairly robust under LED. Therefore, when we marginalize over the reactor angle, we include the external penalty
$\theta_{13} = (8.53\pm 0.13)\degree$~\cite{deSalas:2020pgw}.

In the context of 3+1 neutrinos, the oscillations due to $\Delta m_{41}^2$ are averaged out for the baselines relevant for Daya Bay, RENO and Double Chooz, if $\Delta m_{41}^2 \gtrsim 0.1$ eV$^2$. We focus on these values of $\Delta m_{41}^2$, and hence we can impose a prior on $\theta_{13}$ using the value which has been obtained taking ratios between the near and far detectors of these experiments.

\section{Results}
\label{sec:res}
In this section we present the main results of this paper on the sensitivity of JUNO + TAO to
LED and light sterile neutrinos. In Subsection~\ref{sec:sensitivity_led}
we examine the sensitivity to the LED parameters that can be reached at JUNO+TAO.
In Subsection~\ref{sec:measure_led}
we select a point in the LED parameter space that is consistent with the reactor and Gallium anomalies (in the context of LED) and we show that JUNO+TAO can measure these values of the LED parameters with very good precision.
In Subsection~\ref{sec:3+1}
we present the combined JUNO+TAO sensitivity to 3+1 neutrino oscillation parameters.
Finally, in Subsection~\ref{sec:Distinguishing}
we discuss if JUNO+TAO can distinguish the LED scenario from 3+1 neutrino oscillations.

\subsection{JUNO and TAO sensitivity to large extra dimensions}
\label{sec:sensitivity_led}

We first estimate the sensitivity
of JUNO+TAO
to the LED parameters.
For this purpose,
we create a fake set of data points
without LED oscillations using the following values of the standard neutrino oscillation parameters for normal (NO) and inverted (IO)
neutrino mass ordering~\cite{deSalas:2020pgw}:
\begin{eqnarray}
    \sin^2\theta_{12} &=& 0.318, \quad \sin^2\theta_{13} = 0.022,\quad
    \Delta m_{21}^2 = 7.5\times10^{-5}~\text{eV}^2,\nonumber\\ 
    \Delta m_{31}^{2,\text{NO}} &=& 2.55\times10^{-3}~\text{eV}^2\,,\quad
    \Delta m_{31}^{2,\text{IO}} = -2.45\times10^{-3}~\text{eV}^2\,.
    \label{eq:st_params}
\end{eqnarray}
The analysis of the simulated data is performed by considering the values of the standard oscillation parameters as variables determined by the fit.
The sensitivity to the LED parameters is obtained by marginalizing over the standard oscillation parameters.

Let us comment on the different implications of the normal and inverted orderings for the short-baseline oscillations of reactor neutrinos in the LED framework. For normal ordering, we have that $\lambda_1^{(0)},\lambda_2^{(0)} \ll \lambda_3^{(0)}$, and also $m_1^D, m_2^D \ll m_3^D$. 
Therefore, since for $m_i^D R_{ED} \ll 1$ we have
$ L_i^{0k} \simeq \sqrt{2} m_i^D R_{ED} / k $ for $k\geq1$~\cite{Davoudiasl:2002fq},
the dominant LED contribution to the neutrino oscillation probability comes from the term with $i=3$ in Eq.~\eqref{eq:oscprob}, which is suppressed by the smallness of $|U_{e3}|^2$. If one assumes inverted ordering, however, the dominant LED contributions appear in the terms with $i=1$ and $i=2$, which are not suppressed. Therefore, the LED oscillation effects for fixed $R_{ED}$ and $m_0$ can induce a much larger deviation from standard 3-neutrino oscillations for inverted ordering than for normal ordering. This translates to better sensitivities to the LED parameters for inverted ordering compared to the ones for normal ordering.

\begin{figure}
  \centering
  \includegraphics[width=0.48\textwidth]{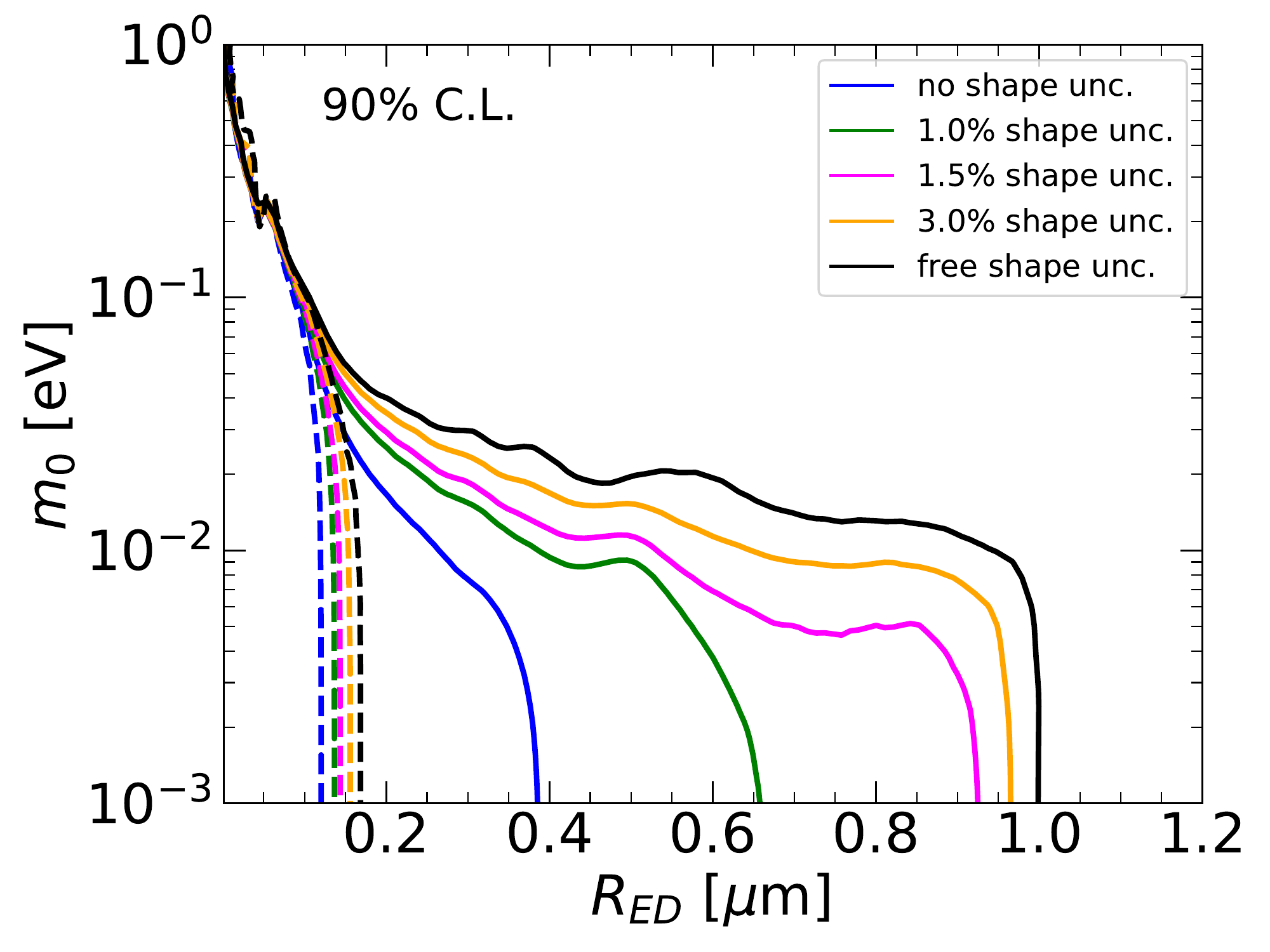}
  \includegraphics[width=0.48\textwidth]{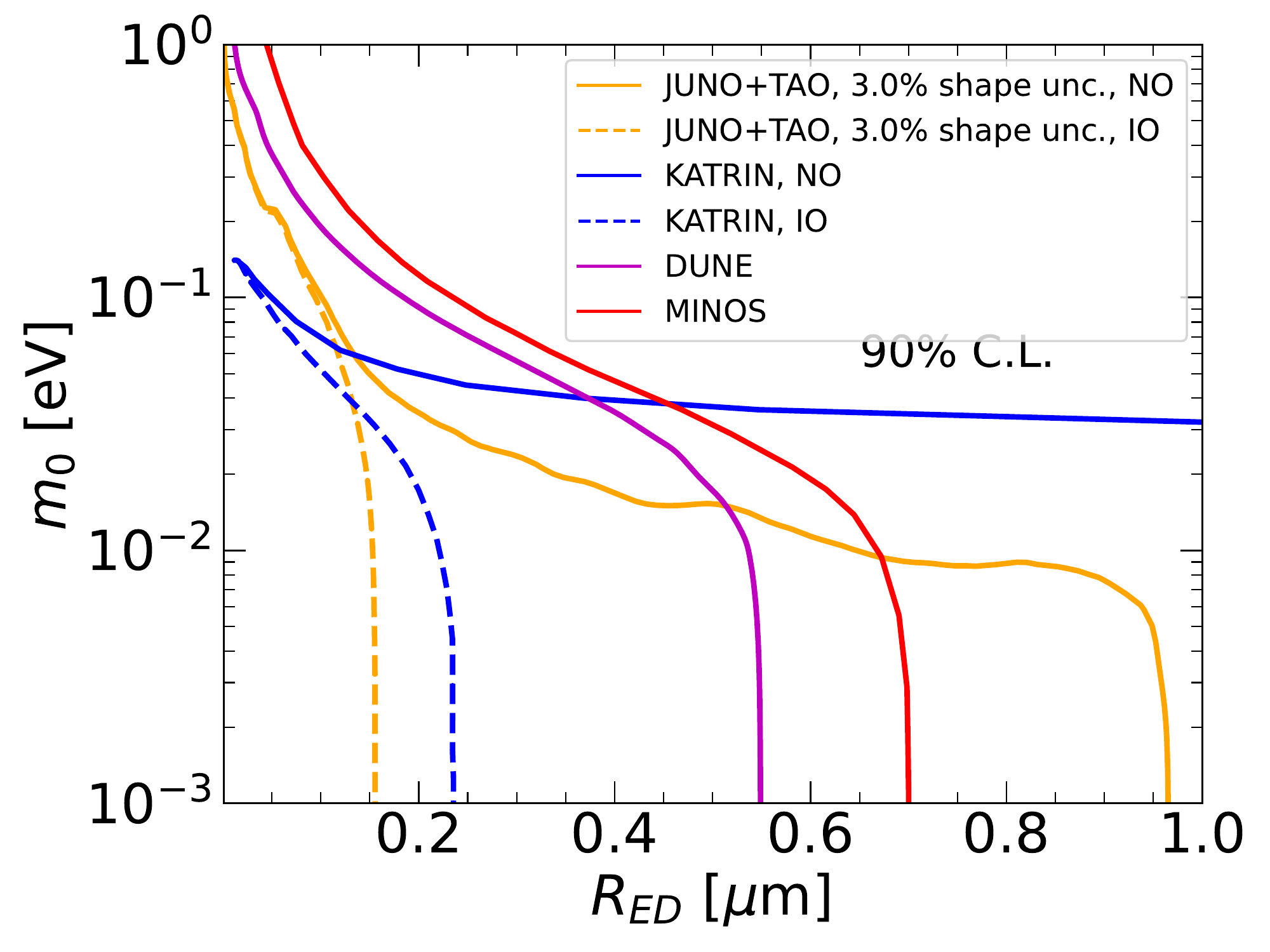}
    \caption{Left: JUNO+TAO 90\% C.L. sensitivity in the $R_{ED}$-$m_0$ plane for several choices of systematic uncertainties.
    The solid and dashed lines correspond, respectively,
    to normal and inverted ordering.
    Right: Comparison of the JUNO+TAO sensitivity with those of KATRIN~\cite{Basto-Gonzalez:2012nel}, DUNE~\cite{Arguelles:2019xgp,DUNE:2020fgq}, and MINOS~\cite{MINOS:2016vvv}.}
  \label{fig:R_m_sens}
\end{figure}

The results for several sets of systematic uncertainties are shown in the left panel of Fig.~\ref{fig:R_m_sens} for normal (solid lines) and inverted ordering (dashed lines). As anticipated, the results for inverted ordering are very strong. Moreover, the result hardly depends on the flux uncertainty. We find that in the case of inverted ordering, JUNO can provide the strongest bound on $R_{ED}$. The bound is stronger than in any current experiment\footnote{Note that experiments measuring $\nu_\mu\to\nu_\mu$ oscillations depend on $|U_{\mu i}|^2$, where all elements are large, and hence the bounds for these experiments are very similar for both orderings.}, and also better than expectations from other future experiments, see right panel of Fig.~\ref{fig:R_m_sens}. The only experiment which can set competitive bounds for inverted ordering is KATRIN. From here on we turn our focus to the discussion of normal ordering.
The blue line in the left panel of Fig.~\ref{fig:R_m_sens} is obtained including the first three systematic uncertainties
in the list in Section~\ref{sec:Statistical} and is the minimal set of uncertainties that we consider in this paper.
We also performed several analyses with different flux shape uncertainties.
This systematic has a very important effect on the sensitivity, as can be seen in the figure.
The green, magenta and orange lines are obtained using a bin-to-bin shape uncertainty of 1\%, 1.5\% and 3\%, respectively. The black line is obtained leaving this systematic uncertainty completely free.

One can see from the left panel of Fig.~\ref{fig:R_m_sens} that $m_0$ is unbounded for very small values of $R_{ED}$.
This is expected, since in the limit $R_{ED}\to0$ the Dirac mass becomes equivalent to the physical neutrino mass and we recover standard neutrino oscillations which do not depend on the overall neutrino mass scale.

Considering the case with free (no) shape uncertainty, at 90\% C.L. for 1 degree of freedom ($\Delta\chi^2 = 2.71$), the LED compactification radius $R_{ED}$ has a sensitivity upper bound of $0.87~\mu\text{m}$ ($0.35~\mu\text{m}$). The bounds for the different systematic uncertainties considered in this section are summarized in Tab.~\ref{tab:bounds} for normal and inverted mass ordering. These results show again the importance of the systematic uncertainties for probing the LED parameters in the case of normal ordering.
Notice however that, for the intermediate cases, the sensitivity to $R_{ED}$ does not significantly change when the shape uncertainty is larger than 3\%. Within this range lies a realistic estimate for the flux-related uncertainties that is of the order of $\sim 5\%$. In what follows, and for comparison purposes, we take as a reference the case with 3\% bin-to-bin uncorrelated uncertainty.

\begin{table}[t!]
\centering
    \addtolength{\leftskip} {-2cm} 
    \addtolength{\rightskip}{-2cm}
  \catcode`?=\active \def?{\hphantom{0}}
   \begin{tabular}{|c|c|c|c|c|c|}
    \hline
    Case & no shape unc. & 1.0\% shape unc. & 1.5\% shape  unc.& 3.0\% shape  unc.& free shape  unc. \\
    \hline
    Bound (90\% C.L.), NO  & 0.35~$\mu$m & 0.57~$\mu$m & 0.65~$\mu$m & 0.81~$\mu$m & 0.87~$\mu$m \\
    \hline
    Bound (90\% C.L.), IO  & 0.115~$\mu$m & 0.128~$\mu$m & 0.134~$\mu$m & 0.142~$\mu$m & 0.151~$\mu$m \\
    \hline
    \end{tabular}
    \caption{JUNO+TAO 90\% C.L. sensitivity upper bound (1 degree of freedom) on the compatification radius $R_{ED}$ for each set of systematic uncertainties considered in Section~\ref{sec:sensitivity_led}.}
    \label{tab:bounds} 
\end{table}

In the right panel of Fig.~\ref{fig:R_m_sens},
we compare the JUNO+TAO sensitivity corresponding to $3\%$ shape uncertainty (orange lines) with the estimated sensitivities of KATRIN~\cite{Basto-Gonzalez:2012nel} (blue lines), DUNE~\cite{Arguelles:2019xgp,DUNE:2020fgq} (purple line; see also Ref.~\cite{Berryman:2016szd}), and with the sensitivity of MINOS~\cite{MINOS:2016vvv} (red line).
Note that the bound obtained by the MINOS collaboration from the data is much stronger than the sensitivity~\cite{MINOS:2016vvv}.
This is due to a fluctuation of the data.
Since we are discussing experimental sensitivities, it is appropriate to compare the JUNO+TAO sensitivity with that of MINOS, not with the MINOS bound obtained from the data.
When the JUNO+TAO data will appear, the resulting bound on the LED parameters will depend on the fluctuation of the data and it will be appropriate to compare it with the MINOS bound obtained from the data.

As can be seen in Fig.~\ref{fig:R_m_sens}, JUNO+TAO can set competitive bounds on the LED parameters, which, depending on the treatment of systematic uncertainties, can be more stringent than those from the other experiments. For the optimistic case of $1\%$ flux-shape systematic uncertainty, the JUNO+TAO sensitivity is compatible with the DUNE sensitivity showing a synergy of future reactor and accelerator neutrino experiments.

\subsection{Measuring LED parameters}
\label{sec:measure_led}

In this subsection we
discuss how well JUNO+TAO can measure values of the LED parameters
that are consistent with a LED explanation of the reactor and Gallium anomalies~\cite{Machado:2011kt}.
For this purpose, we generate a simulated data set with the same values of the
standard oscillation parameters (focusing here only on normal ordering) considered in the previous subsection
(see Eq.~\eqref{eq:st_params})
and the following specific values of the LED parameters:
\begin{eqnarray}
    R_{ED} &=& 0.5~\mu\text{m}\,,
    \\
    m_0 &=& 0.03~\text{eV}\,.
\label{eq:benchmark}
\end{eqnarray}
We assume the neutrino mass ordering to be normal. Using this benchmark point one can explain the reactor antineutrino anomaly~\cite{Mention:2011rk}
and the Gallium neutrino anomaly~\cite{Abdurashitov:2005tb,Laveder:2007zz,Giunti:2006bj,Barinov:2021asz}
as a result of LED oscillations~\cite{Machado:2011kt}.

\begin{figure}
  \centering
  \includegraphics[width=0.6\textwidth]{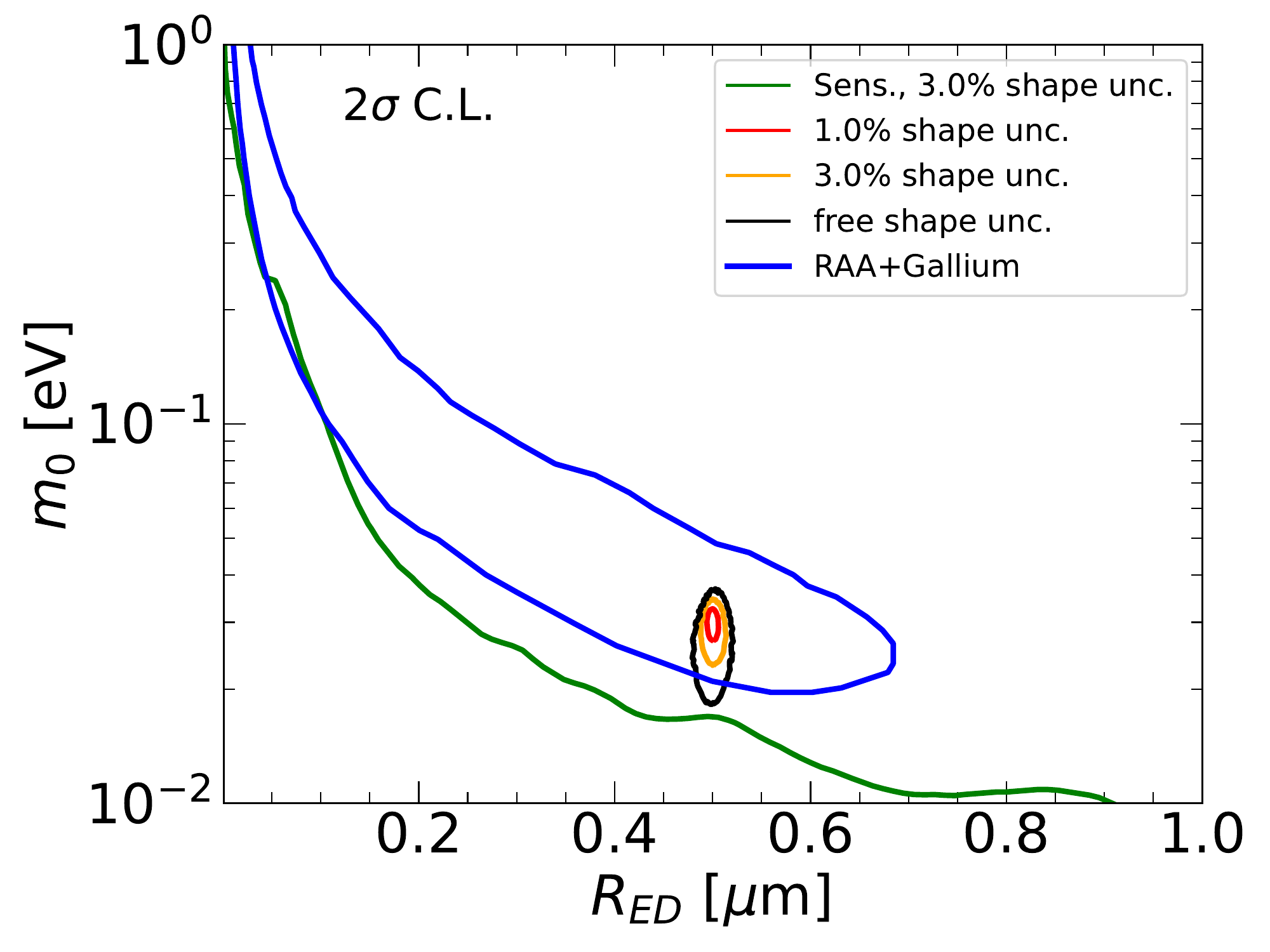}
    \caption{2$\sigma$ regions obtained from the analysis of fake data in presence of LED oscillations for a several assumptions on the flux shape uncertainty (red, orange and black lines), in comparison with the region allowed from reactor and Gallium anomalies (blue). We also show the JUNO+TAO sensitivity (green line) to emphasize that JUNO+TAO can test all of the values preferred by the reactor and Gallium anomalies.}
  \label{fig:R_m_dat}
\end{figure}

As can be seen in Fig.~\ref{fig:R_m_dat}, relatively small closed contours at 2$\sigma$ can be obtained with this benchmark point. Indeed, the expected sensitivity to measure the LED parameters is at a similar level ($\lesssim 1\%$) as the expected sensitivity to the standard neutrino oscillation parameters~\cite{JUNO:2021vlw}.
The JUNO+TAO sensitivity with $3\%$ shape uncertainty (full green line) is also shown in Fig.~\ref{fig:R_m_dat} in comparison with the allowed region from the reactor and Gallium anomalies, extracted from Ref.~\cite{Machado:2011kt} (blue line). 

We considered for illustration only one point in the region of the LED parameters
inside of the blue contour in Fig.~\ref{fig:R_m_dat}, that is motivated by the LED explanation of the reactor and Gallium anomalies~\cite{Machado:2011kt}.
However, it should be noted that basically all of the region inside of the blue contour lies within the sensitivity reach of JUNO+TAO. Therefore, it is plausible that a similar precision of the JUNO+TAO determination of the LED parameters can be obtained for any benchmark point compatible with the LED explanation of the reactor and Gallium anomalies. 

\subsection{Light sterile neutrinos at JUNO and TAO}
\label{sec:3+1}

The JUNO and TAO detectors can also be used to measure or bound active-sterile neutrino oscillation parameters~\cite{JUNO:2021vlw,Berryman:2021xsi}. In this subsection we explore the sensitivity to $\sin^22\theta_{14}$ and $\Delta m_{41}^2$
in the 3+1 neutrino mixing framework. The analysis follows the description given above, with the only difference that now the LED neutrino oscillation probability is replaced by the oscillation probability describing 3+1 neutrino mixing.

\begin{figure}
  \centering
  \includegraphics[width=0.49\textwidth]{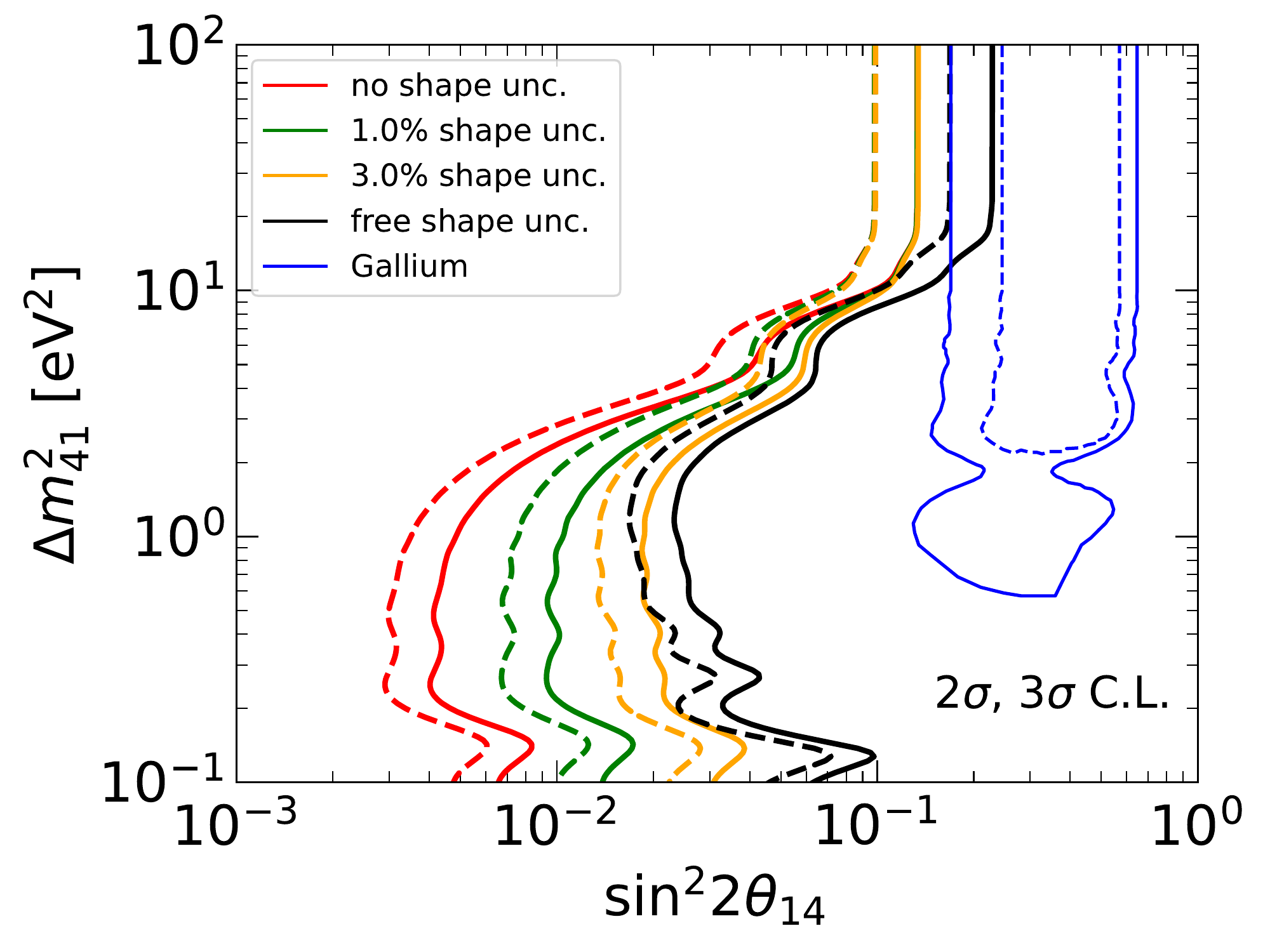}
  \includegraphics[width=0.49\textwidth]{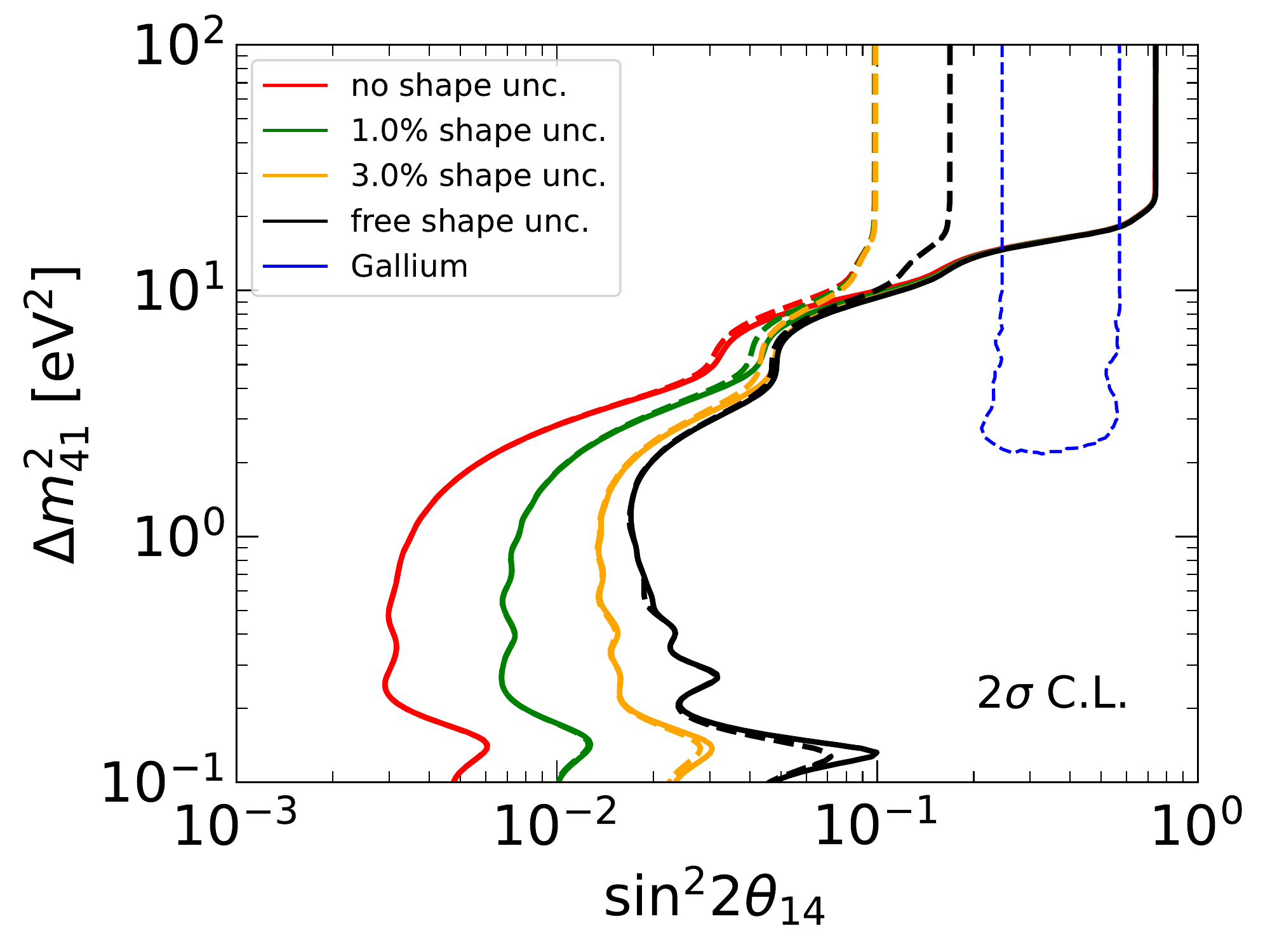}
    \caption{Left: The 2$\sigma$ (dashed) and 3$\sigma$ (solid) sensitivity to sterile oscillation parameters obtained in the analysis of JUNO+TAO. We also show the region preferred by Gallium experiments in blue~\cite{Barinov:2021mjj}. Right: Impact of the overall flux normalization on the sensitivity. The dashed lines correspond to the 2$\sigma$ sensitivity curves in the left panel, while the solid lines are obtained leaving the flux normalization free.}
  \label{fig:sterile_sens}
\end{figure}

The results of our analyses are shown in the left panel of Fig.~\ref{fig:sterile_sens}. We show the sensitivity expected at 2$\sigma$ (dashed) and 3$\sigma$ (solid) confidence level for different
choices of the flux shape uncertainty,
as we have done in Section~\ref{sec:sensitivity_led} for LED-induced oscillations.

Note that our results on the sensitivity differ from those obtained in Ref.~\cite{Berryman:2021xsi}, particularly for very small and large values of $\Delta m_{41}^2$. This is due to the fact that there are crucial differences in the analyses. The total exposure assumed in this paper is larger than that assumed in Ref.~\cite{Berryman:2021xsi}.
On the other hand, our treatment of backgrounds is more conservative. Most importantly, the authors of Ref.~\cite{Berryman:2021xsi} studied the TAO sensitivity alone, without taking into account the contribution from the far detector.
The same argument applies when comparing with the results obtained by the TAO collaboration in Ref.~\cite{JUNO:2020ijm}. The addition of the far detector allows to bound the mixing also for small and large values of $\Delta m_{41}^2$. It should be also noted, that the treatment of the shape uncertainty in Ref.~\cite{JUNO:2020ijm} slightly differs from ours.

In the left panel of Fig.~\ref{fig:sterile_sens},
we also show the preferred region obtained from the combined analysis of the Gallium experiments~\cite{Barinov:2021mjj}, where of particular importance are the recent results of the BEST experiment~\cite{Barinov:2021mjj}. We find that JUNO+TAO can exclude nearly all of the parameter space preferred by the Gallium experiments. Indeed, only for the shape-free analysis there is some overlap at 3$\sigma$ between the Gallium and JUNO+TAO regions for $\Delta m_{41}^2 \gtrsim 10 \, \text{eV}^2$.
Note that most of the parameter space preferred from the analysis of Neutrino-4 data~\cite{Serebrov:2020kmd,Giunti:2021iti}, which is in agreement with the BEST results, could be therefore also ruled out by JUNO+TAO. 
Also, a large part of the parameter space allowed by the analysis in Ref.~\cite{Denton:2021czb} of the MicroBooNE data~\cite{MicroBooNE:2021sne,MicroBooNE:2021jwr,MicroBooNE:2021rmx} in terms of $\nu_e$ disappearance is within the sensitivity reach of JUNO+TAO.

It must be mentioned that the bound on $\sin^22\theta_{14}$ at large values of $\Delta m_{41}^2$ is due to the fact that we assume a 2\% total normalization uncertainty on the flux. If we relax this assumption and leave the normalization free, we obtain the solid lines in the right panel of Fig.~\ref{fig:sterile_sens}, where we also show (dashed lines) the bounds using our nominal 2\% penalty (these are the same as in the left panel). Note that even with a free normalization and free shape uncertainty JUNO+TAO can test the Gallium region for $\Delta m_{41}^2\lesssim10-20$~eV$^2$.

\subsection{Distinguishing LED oscillations from light sterile neutrinos}
\label{sec:Distinguishing}

\begin{figure}
  \centering
  \includegraphics[width=0.49\textwidth]{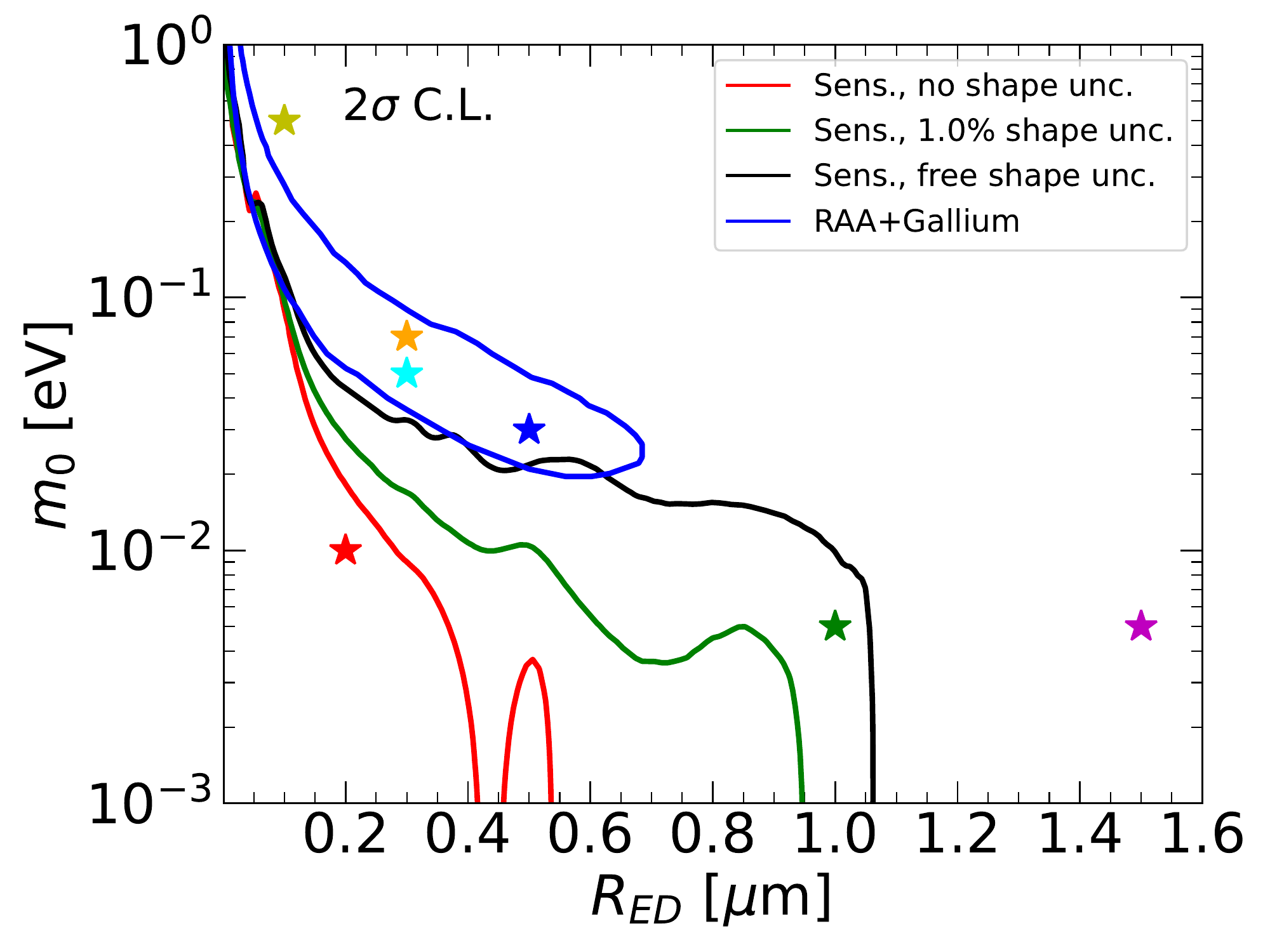}
  \includegraphics[width=0.49\textwidth]{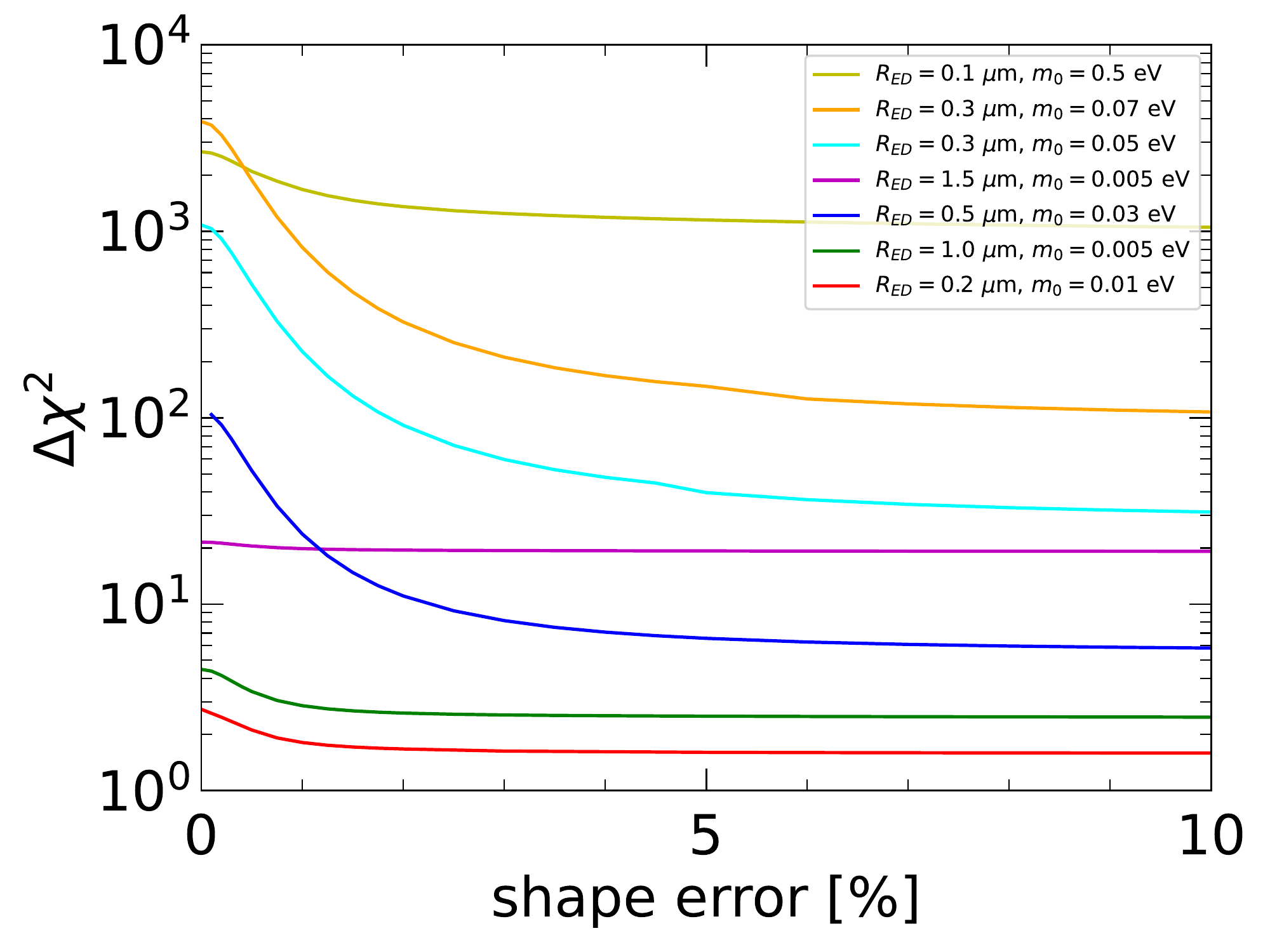}
    \caption{Left: Positions of the benchmark points (the stars) chosen to generate the fake data, in comparison with the JUNO+TAO sensitivity and the preferred region from Gallium and reactor experiments. Right: The minimum $\Delta \chi^2$ value obtained when analyzing the LED generated fake data using 3+1 neutrino oscillation parameters for several LED data sets (as indicated in the legend).}
  \label{fig:sterile_dat}
\end{figure}

Finally, in this subsection we investigate if JUNO+TAO can distinguish LED oscillations from 3+1 oscillations. For this purpose, we generate fake data sets consistent with LED oscillations. We consider several benchmark points in the LED parameter space, summarized in the first column of Table~\ref{tab:benchmarks} and graphically indicated in the left panel of Fig.~\ref{fig:sterile_dat}. 
We take points inside and outside of the sensitivity regions of JUNO+TAO and some points inside the region preferred by the LED explanation of the reactor and Gallium anomalies~\cite{Machado:2011kt}. In the right panel of Fig.~\ref{fig:sterile_dat}, we plot the minimal $\Delta\chi^2$ value obtained for each analysis as a function of the shape uncertainty. 
As one can see, the effect of the shape uncertainty saturates for values between approximately 3\% and 8\%, depending on the benchmark. We have checked that even for larger values of the shape uncertainty no significant change of $\Delta\chi^2$ can be observed. This is a result of our two detector analysis. Since the flux uncertainties are correlated among detectors (and detector segments) the effect of the flux uncertainty saturates at some point, and increasing the uncertainty has no further effect anymore.
In the last two columns of Table~\ref{tab:benchmarks} we report the values of the 3+1 oscillation parameters for which the minimal $\Delta\chi^2$ has been found in the case of 10\% shape uncertainty.
Interestingly, we find that the minimal $\Delta\chi^2$ values span several orders of magnitude, as shown by the second column of Table~\ref{tab:benchmarks} and the right panel of Fig.~\ref{fig:sterile_dat}. 

We find that $\Delta\chi^2_{\text{min}}$ is very small in the cases of the red and green benchmark points.
The red point lies outside of the sensitivity region of JUNO+TAO,
indicated in the left panel of Fig.~\ref{fig:sterile_dat}
by the red, green, and black lines for negligible, 1\%, and free flux shape uncertainties,
respectively (see the discussion in Section~\ref{sec:sensitivity_led}).
Therefore, in this case the simulated data are not very different from those given by standard 3-neutrino oscillations.
Indeed, they are best fitted within the 3+1 scenario with small values of $\sin^22\theta_{14}$.
The green point lies within the sensitivity region of JUNO+TAO if the shape uncertainties are negligible or small (1\%), but it is outside of the sensitivity region for free shape uncertainties.
Therefore, in this case the data are not fitted
well with 3-neutrino oscillations. However, we find that 3+1 oscillations provide a decent fit with $\Delta\chi^2_{\text{min}}<5$.
As a result, for these benchmark points a distinction of the two models is not possible.

For the blue benchmark point (note that it corresponds to the benchmark point of Section~\ref{sec:measure_led}), we find that $\Delta\chi^2_{\text{min}}$ depends importantly on the value of the shape uncertainty. If the shape is mostly unknown we can fit the LED data rather well using 3+1 oscillations and the $\Delta\chi^2_{\text{min}}$ is small. However, for shape uncertainties smaller than about 2\% we obtain $\Delta\chi_{\text{min}}^2\gtrsim15$ and a distinction between the two models should not be possible.

For the other benchmark points the fit of the simulated LED data in terms of 3+1 oscillations is rather poor, even for a conservative choice of the shape uncertainty.
This means that in these cases a distinction between the two models is feasible with JUNO+TAO. In general, we find that if the benchmark point is far away from the sensitivity curves a better distinction is possible. The reason for this behavior is due to the fact that the amplitudes $L_i^{0n}$ increase when increasing $m_i^D$ and $R_{ED}$. Therefore, if we generate the fake data with points which are far away from the sensitivity curve, more oscillation lengths become relevant, and the data can not be fitted well anymore within the 3+1 scenario. A better fit could be performed considering $N > 1$ sterile neutrinos, because of the equivalence of the LED and $3+N$ sterile neutrino models discussed in Ref.~\cite{Esmaili:2014esa}.

Note that we did not assign a precise statistical meaning to the $\Delta\chi^2$'s that we obtained, because a precise quantitative statistical comparison of the LED and 3+1 models would require a complex calculation of the distribution of the $\Delta\chi^2$ in the frequentist approach,
or a complex evaluation of the Bayes factor in the Bayesian approach.
As an indicative value of the $\Delta\chi^2$ that may allow to distinguish the two models, one can consider $\Delta\chi^2=9$, that corresponds to $3\sigma$ for a $\chi^2$ distribution with one degree of freedom.
However, in the case of a real observation of an anomalous result at JUNO+TAO a more sophisticated statistical estimate should be done.

\begin{table}[t!]
\centering
  \catcode`?=\active \def?{\hphantom{0}}
   \begin{tabular}{|c|c|c|c|}
    \hline
    Benchmark point & $\Delta \chi_{\text{min}}^2$  & $\sin^22\theta_{14}$ & $\Delta m_{41}^2$ \\
    \hline
    $R_{ED}=0.1~\mu$m, $m_0=0.5$~eV     &    1050   &   0.252   &    0.137~eV$^2$    \\
    $R_{ED}=0.3~\mu$m, $m_0=0.07$~eV    &    107    &   0.092   &    0.036~eV$^2$    \\
    $R_{ED}=0.3~\mu$m, $m_0=0.05$~eV    &    31     &   0.047   &    0.034~eV$^2$    \\
    $R_{ED}=1.5~\mu$m, $m_0=0.005$~eV   &    19     &   0.025   &    0.116~eV$^2$    \\
    $R_{ED}=0.5~\mu$m, $m_0=0.03$~eV    &    5.8    &   0.059   &    0.155~eV$^2$    \\
    $R_{ED}=1.0~\mu$m, $m_0=0.005$~eV   &    2.5    &   0.010   &    0.115~eV$^2$    \\
    $R_{ED}=0.2~\mu$m, $m_0=0.01$~eV    &    1.6    &   0.001   &    0.208~eV$^2$    \\
    \hline
    \end{tabular}
    \caption{The benchmark points used to generate the fake data sets and the corresponding minimal $\Delta\chi^2$ values and values of the 3+1 oscillation parameters at the minimum for a shape uncertainty of 10\%.}
    \label{tab:benchmarks} 
\end{table}

\section{Conclusions}
\label{sec:conc}

We have discussed the JUNO and TAO sensitivity to neutrino oscillations
induced by Large Extra Dimensions and by 3+1 active-sterile neutrino mixing.
We have shown that the experiments can set competitive bounds on the parameters for both scenarios.

In the LED scenario, and for the realistic case of $3\%$~flux-shape systematic uncertainty, we found that JUNO+TAO is sensitive to values of the compactification radius $R_{ED}>0.81\,\mu \text{m}$ at $90\%$ C.L. (for NO), which is in the sensitivity range of LBL experiments such as MINOS/MINOS+ and DUNE.

We have shown that, in the case of light sterile neutrinos, JUNO+TAO can test most of the region preferred by the analysis of the data of the Gallium experiments that includes the recent results of the BEST experiment~\cite{Barinov:2021mjj}.

Considering the LED framework,
since there is no updated analysis of the Gallium data that includes the BEST data, we considered the results in Ref.~\cite{Machado:2011kt} of the LED explanation of the reactor anomaly and the GALLEX+SAGE Gallium anomaly,
that is anyhow consistent with the BEST data.
We have shown that JUNO+TAO can test all the $2\sigma$ region of the LED parameter space that is allowed by the analysis in Ref.~\cite{Machado:2011kt}.
Moreover, we have shown that for a benchmark point inside of this allowed region, JUNO+TAO can measure the LED parameters with a precision that is similar to that of the standard neutrino oscillation parameters.
It is plausible that a similar precision can be obtained for any value of the LED parameters that lies in the $2\sigma$ region allowed by the analysis in Ref.~\cite{Machado:2011kt}, since it is within the sensitivity reach of JUNO+TAO.

We have discussed also the possibility to distinguish LED oscillations from 3+1 active-sterile oscillations in case of a deviation of the JUNO+TAO data from the prediction of three-neutrino mixing. We have shown that the possibility of a distinction of the two models depends on the knowledge of the reactor flux shape that will be reached when JUNO and TAO will operate and on the true values of the LED parameters. In general,
it is likely that the two models can be distinguished
if the true values of the LED parameters lie in the sensitivity region of JUNO+TAO.

It is worth mentioning that also other types of experiments and analyses have been performed in order to test the existence of Large Extra Dimensions, as for instance tabletop gravitational experiments~\cite{Long:1998dk,Krause:1999ry,Fischbach:2001ry,Adelberger:2002ic},
collider experiments~\cite{Rizzo:1998fm,Hewett:1998sn,D0:2000cve,DELPHI:2000ztm,DELPHI:2008uka},
analyses of astrophysical~\cite{Cullen:1999hc,Barger:1999jf,Hanhart:2000er,Hannestad:2001jv,Hannestad:2003yd} and cosmological data~\cite{Hall:1999mk,Hannestad:2001nq,Fairbairn:2001ct}. In the case of astrophysical data, strong constraints ranging in $R_{ED}<0.16-916\,\text{nm}$ can be obtained. However, these limits depend on the technique and some assumptions~\cite{ParticleDataGroup:2020ssz}.
Interestingly, neutrino oscillation experiments provide a stronger constraint than tabletop experiments, which test the gravitational force at sub-millimeter distances producing a limit of $R_{ED}<37~\mu \text{m}$ at 95\% of C.L~\cite{ParticleDataGroup:2020ssz}.

Also in the case of light sterile neutrinos other searches exist, for example using cosmological probes or $\beta$-decay (see the recent reviews in Ref.~\cite{Boser:2019rta,Hagstotz:2020ukm,Dasgupta:2021ies}). 
In all cases, neutrino oscillation experiments provide complementary tests of these scenarios and an observation in one sector and non-observation in another could have interesting impacts on our understanding of these fields.

\section*{Acknowledgments}
CG and CAT are supported by the research grant ``The Dark Universe: A Synergic Multimessenger Approach'' number 2017X7X85K under the program ``PRIN 2017'' funded by the Ministero dell'Istruzione, Universit\`a e della Ricerca (MIUR). AAQ thanks Arman Esmaili and Hiroshi Nunokawa for useful discussions regarding the LED model. VSBG's work was partly supported by CNPQ under grant No. 151087/2014-8 and Universidad de Pamplona through Proyectos Convocatoria Interna 2018.


%

\end{document}